\documentclass[aps,pra,superscriptaddress,twocolumn,nofootinbib]{revtex4-1}
\usepackage{amsmath}
\usepackage{amssymb}
\usepackage{dsfont}
\usepackage{color}
\makeatletter
\usepackage{amssymb}
\usepackage{graphicx}
\usepackage{epstopdf}
\usepackage{epsfig}
\usepackage{subfigure}
\usepackage{xfrac}
\usepackage[colorlinks,urlcolor=blue,citecolor=blue]{hyperref}
\makeatother

\newcommand{\ket}[1]{|#1\rangle}
\newcommand{\bra}[1]{\langle#1|}

\newcommand{\C}{\hat{\psi}^{\dagger}}
\newcommand{\D}{\hat{\psi}}

\newcommand{\A}[1]{\hat{a}_{\bf #1}}
\newcommand{\AC}[1]{\hat{a}_{\bf #1}^{\dagger}}

\usepackage{cancel,ifthen}
\usepackage[normalem]{ulem}

\newcommand{\bref}[3]{{#1}\href{#2}{#3}}

\newcommand{\comment}[2][NoInPuT]{\ifthenelse{\equal{#1}{NoInPuT}}{}{{\color{blue}\sout{#1}}}{\color{red} #2}}
\begin{document}

\title{Quantum droplets in a beyond-mean-field density-dependent gauge theory}

\author{Matthew Edmonds}
\email{m.edmonds@uq.edu.au}
\affiliation{ARC Centre of Excellence in Future Low-Energy Electronics Technologies,
School of Mathematics and Physics, University of Queensland, St Lucia, QLD 4072, Australia}
\affiliation{Department of Physics \& Research and Education Center for Natural Sciences, Keio University, Hiyoshi 4-1-1, Yokohama, Kanagawa 223-8521, Japan}
\author{Patrik \"Ohberg}
\affiliation{Institute of Photonics and Quantum Sciences, SUPA, Heriot-Watt University, Edinburgh EH14 4AS, UK}

\date{\today{}}

\begin{abstract}\noindent
The beyond-mean-field corrections appropriate to a bosonic many-body system experiencing a density-dependent gauge potential are derived, and from this the dimensional hierarchy of quantum droplet solutions are explored. Non-stationary quantum droplet solutions are supported by a single interaction parameter characterising the strength of the gauge potential, while in one dimension the beyond-mean-field theory can be solved exactly to yield chiral quantum droplets and dark soliton-like excitations. Numerical simulations of single and pairs of chiral droplets indicate a rich dynamics in the beyond-mean-field regime. 
\end{abstract}
\maketitle
{\it Introduction. }
Interactions underpin the emergence of a wide variety of quantum mechanical phenomena in condensed matter systems. Here, the particulars of the atoms statistics, dimensionality and potential environment contribute to the overall equilibrium description of the many-particle state. Prominent illustrations are given by superconductivity \cite{bruus_book} and the family of quantum Hall effects \cite{klitzing_2020} in fermionic systems, while for bosonic systems the Lieb-Liniger \cite{franchini_book,cazalilla_2011} and closely related Tonks-Girardeau models \cite{sowinski_2019} describe the many-body physics in this limit, possessing solutions which allow important insight into weakly and strongly coupled interacting many-body quantum gases \cite{mistakidis_2023}. Closely related to this, the BCS to BEC crossover remains an important concept for understanding the interplay of quantum statistics and many-body interactions \cite{ohashi_2020,strinati_2018}.

Over the last few years systems comprised from ensembles of bosonic particles  possessing either magnetic dipole-dipole interactions \cite{kadau_2016,schmitt_2016,barbut_2016} or formed from binary mixtures \cite{cabrera_2018,semeghini_2018,cheiney_2018,ferioli_2019,derrico_2019,cavicchioli_2025} have received an intense amount of theoretical and experimental attention, due to the discovery and subsequent explanation of their ability to host droplet-like phases -- where the weakly interacting system can exist in regimes where established theories predicted the collapse of the mean-field wave function. The stability of the liquid-like phase has been accounted for by beyond-mean-field corrections \cite{lima_2011,lima_2012,petrov_2015} to the ground state energy of the system in the form of the Lee-Huang-Yang (LHY) term \cite{lee_1957}. 

Quantum droplet phases now represent an established paradigm within the quantum gas community. Studies have also focused on realizing pure LHY fluids  \cite{jorgensen_2018,skov_2021}, extensions to coherently coupled systems \cite{chiquillo_2019}, spinor systems \cite{uchino_2010,yogurt_2023}, Bose-Fermi mixtures \cite{adhikari_2018,rakshit_2019} and understanding the fundamental physics of dipolar droplets in terms of the quantum fluctuations of the gas \cite{edler_2017}. Recently the enigmatic supersolid phase where off-diagonal and long-range order can paradoxically coexist was demonstrated in dipolar condensates \cite{tanzi_2019,bottcher_2019,chomaz_2019}. Parallel to investigations in quantum gas systems, droplet-like phases of matter have also been studied in the Helium fluids \cite{barranco_2006}, optomechanical \cite{walker_2022}, semiconductor \cite{hunter_2014} and optical systems realising photonic droplets \cite{wilson_2018}.

Ultracold quantum gases represent highly amenable platforms for simulating synthetic forms of matter, close to absolute zero where quantum mechanical effects can be clearly studied and accurately modelled \cite{schafer_2020}. Engineering synthetic degrees of freedom has allowed a series of state-of-the-art experiments to simulate orbital magnetism \cite{lin_2009,lin_2009b,lin_2011} with quantum gases. While these accomplishments brought important effects from condensed matter into the cold atom realm such as spin-orbit coupling \cite{lin_2011b}, spin-Hall phenomena \cite{beeler_2013} and synthetic dimensions \cite{celi_2014}; a fundamental limitation existed in that these systems describe static gauge theories such that the gauge potential is not itself a dynamical object.

The last few years has seen the experimental demonstration of quasi-dynamical gauge theories where the synthetic gauge degree of freedom is coupled to the generally time-dependent quantum state of the system, thus providing a form of nonlinear feedback between the gauge and matter degrees of freedom. Pioneering experiments realised density-dependent gauge theories in optical lattice geometries for both bosons \cite{clark_2018,kwan_2024} and fermions \cite{gorg_2019} and also for an ensemble of Rydberg atoms \cite{lienhard_2020}. Following this two groups independently realised density-dependent gauge potentials in the continuum where a domain wall excitation was observed \cite{yao_2022}, and a type of topological gauge theory associated with the  quantum Hall effect \cite{aglietti_1996} was realised \cite{frolian_2022} possessing an unusual type of edge mode in the form of a chiral soliton \cite{chisholm_2022}.

These synthetic forms of quasi-dynamical gauge theories have stimulated further questions regarding both the inherent phenomenology of these unusual systems; such as the nature of their nonlinear \cite{dingwall_2019,bhat_2021,jia_2022,chen_2022,zhang_2023,gao_2023,xu_2023,arazo_2023,faugno_2024}, spinor \cite{zezyulin_2018,xu_2021,abdullaev_2023,arazo_2024} and superfluid states \cite{bhat_2023,bhat_2024,gao_2024} as well as more fundamental goals concerning the route to simulating a truly dynamical gauge field with quantum gas systems \cite{rojas_2020,rojas_2023}. While existing theories account for density-dependent magnetism in the dilute weakly interacting limit, given the interest in beyond-mean-field phases, there is currently a gap in describing the phenomenology of this inherently many-body system in a regime where LHY physics comes into play. The purpose of this work is to derive such a theory appropriate to a system of many-body interacting bosons, and from this explore the allowed quantum droplet solutions across the parameter space of the model.

{\it Beyond-mean-field model. } We consider the following second-quantized Hamiltonian describing a system of $N$ interacting bosons 
\begin{equation}\label{eqn:ham_zero}
    \hat{H}=\int d{\bf r}\ \C \bigg[\frac{1}{2m}\bigg(\hat{\bf p}-\hat{A}({\bf r})\bigg)^2+\frac{g_d}{2}\C\D \bigg]\D
\end{equation}
here $\D\equiv\hat{\psi}({\bf r})$ defines the annihilation operator for a boson at position ${\bf r}$, while the gauge potential is defined as $\hat{A}={\bf a}\C({\bf r})\D({\bf r})$ where {\bf a} defines the vectorial strength of the density-dependent gauge potential while $g_d$ defines the dimensionally-dependent strength of the $s$-wave interactions. 
Then the Heisenberg equation for $\hat{\psi}({\bf r})$ is 
\begin{equation}\label{eqn:hem}
    i\hbar\frac{\partial}{\partial t}\D({\bf r},t)=\bigg[\frac{1}{2m}\bigg(\hat{\bf p}-\hat{A}({\bf r})\bigg)^2+{\bf a}\cdot\hat{J}+g_d\C\D\bigg]\D,
\end{equation}
where the second-quantized current operator $\hat{J}\equiv\hat{J}({\bf r})$ is defined as
\begin{equation}\label{eqn:co}
    \hat{J}=\frac{\hbar}{2mi}\bigg[\bigg(\nabla+\frac{i}{\hbar}\hat{A}({\bf r})\bigg)\C\D-\C\bigg(\nabla-\frac{i}{\hbar}\hat{A}({\bf r})\bigg)\D\bigg].
\end{equation}
We transform Eq.~\eqref{eqn:ham_zero} into momentum space which gives the total Hamiltonian $\hat{H}=\hat{H}_{1}+\hat{H}_{2}+\hat{H}_{3}$ where 
\begin{subequations}\label{eqn:ham_ft}
\begin{align}
    \hat{H}_{1}&=\sum_{\bf k}\epsilon_{\bf k}^{0}\AC{k}\A{k},\\
    \hat{H}_{2}&{=}\frac{1}{2\mathcal{V}}\sum_{{\bf k}{\bf k}'{\bf q}}\bigg\{g_{d}{-}\frac{\hbar{\bf a}}{m}\cdot\big(2{\bf k}{+}{\bf q}\big)\bigg\}\hat{a}^{\dagger}_{{\bf k}+{\bf q}}\hat{a}^{\dagger}_{{\bf k}'-{\bf q}}\A{k'}\A{k},\\
    \hat{H}_{3}&=\frac{1}{\mathcal{V}^2}\frac{a^2}{2m}\sum_{\substack{{\bf k}{\bf k}'{\bf k}''\\ {\bf q}{\bf q}'}}\hat{a}^{\dagger}_{{\bf k}+{\bf q}}\hat{a}^{\dagger}_{{\bf k}'-{\bf q}+{\bf q}'}\hat{a}^{\dagger}_{{\bf k}''-{\bf q}'}\hat{a}_{{\bf k}''}\hat{a}_{{\bf k}'}\hat{a}_{{\bf k}}.
\end{align}
\end{subequations}
Here $\mathcal{V}$ defines the volume of the system and $\epsilon_{\bf k}^{0}=\hbar^2{\bf k}^2/2m$ is the single-particle energy. The terms $\hat{H}_1$, $\hat{H}_2$, and $\hat{H}_3$ contribute respectively the single, two and effective three-body interactions to the total Hamiltonian $\hat{H}$. Applying the Bogoliubov approximation \cite{pethick_book} to the system \eqref{eqn:ham_ft} and diagonalizing the resulting quadratic Hamiltonian leads to the expression
\begin{align}\nonumber
    \hat{H}_{\rm Bog.}&=\frac{g_dn_0N}{2}+\frac{a^2n_{0}^2N}{2m}+\sum_{{\bf k}\neq 0}\bigg\{\mathcal{E}_{\bf k}+\frac{2\hbar n_0}{m}{\bf a}\cdot{\bf k}\bigg\}\hat{\alpha}_{\bf k}^{\dagger}\hat{\alpha}_{\bf k}
    \\ \label{eqn:ham_diag}&-\frac{1}{2}\sum_{{\bf k}\neq 0}\bigg\{\epsilon_{\bf k}^{0}+n_0g_d+\frac{3a^2n_{0}^2}{m}-\mathcal{E}_{\bf k}\bigg\},
\end{align}
where $\mathcal{E}_{\bf k}=\left[\epsilon_{\bf k}^{0}\big(\epsilon_{\bf k}^{0}+2n_0g_d+6a^2n_{0}^{2}/m)\right]^{1/2}$. Equation \eqref{eqn:ham_diag} introduces the canonical quasi-particle operator $\hat{\alpha}_{\bf k}$ defined as $\hat{a}_{\bf k}=u_k\hat{\alpha}_{\bf k}-v_k\hat{\alpha}^{\dagger}_{-{\bf k}}$ satisfying $\big[\hat{\alpha}_{\bf k},\hat{\alpha}^{\dagger}_{{\bf k}'}\big]=\delta_{{\bf k},{\bf k}'}$, $n_0=N/\mathcal{V}$ defines the homogeneous density, while the two amplitude functions are defined by $u_{k}^{2}=\frac{1}{2}\left(\xi_{\bf k}/\mathcal{E}_{\bf k}+1\right)$ and $v_{k}^2=\left(\xi_{\bf k}/\mathcal{E}_{\bf k}-1\right)$, where $\xi_{\bf k}=\epsilon_{\bf k}^{0}+n_0g_d+3a^2n_{0}^{2}/m$ is the difference between the Hartree-Fock energy of a particle and the chemical potential $\mu_0=n_0g_d+3a^2n_{0}^{2}/m$. The diagonalized Hamiltonian's eigenfunctions from Eq.~\eqref{eqn:ham_diag} are given by Fock states $\ket{n_{{\bf k}_1},n_{{\bf k}_2},\dots}$ where the momentum mode ${\bf k}_1$ possesses $n_{{\bf k}_1}$ Bogolon quasiparticles. The eigenvalue equation for the quasiparticle number operator $\hat{\alpha}^{\dagger}_{{\bf k}_{j}}\hat{\alpha}_{{\bf k}_j}$ is 
\begin{equation}\label{eqn:bnum}
    \hat{\alpha}^{\dagger}_{{\bf k}_{j}}\hat{\alpha}_{{\bf k}_{j}}\ket{\dots,n_{{\bf k}_j},\dots}=n_{{\bf k}_j}\ket{\dots,n_{{\bf k}_j},\dots}
\end{equation}
and the Bogoliubov ground state associated with Eq.~\eqref{eqn:bnum} is $\ket{\psi_{\rm Bog.}^{0}}=\ket{0\dots 0}$ \cite{ueda_book}. Corrections to the ground state energy can then be obtained from Eq.~\eqref{eqn:ham_diag} from $E_{\rm gnd}^{(d)}=\bra{\psi^{0}_{\rm Bog.}}\hat{H}_{\rm Bog.}\ket{\psi^{0}_{\rm Bog.}}$. The energy $E_{\rm gnd}^{(d)}$ is divergent in $d\geq 2$, requiring regularization for both the two and three-dimensional cases. Defining $E_{\rm gnd.}^{(d)}/L^d=a^2n_{\rm d}^{3}/2m+g_{d}n_{d}^2\mathcal{F}_{\rm reg.}^{(d)}/2$, one obtains
\begin{subequations}\label{eqn:freg}
    \begin{align} \label{eqn:f_3d}
        \mathcal{F}_{\rm reg}^{(3)}&{=}1{+}\frac{128}{15\sqrt{\pi}}\big(n_{3} a_{s}^{3}\big)^{1/2}\bigg[1{+}\frac{3a^2 n_{3}}{mg_3}\bigg]^{5/2},\\ \label{eqn:f_2d}
        \mathcal{F}_{\rm reg}^{(2)}&{=}1{+}\frac{g_2m}{2\pi\hbar^2}\ln\bigg(\frac{n_{2}g_2{+}3a^2n_{2}^{2}/m}{\Delta_{\rm cut}}\bigg)\bigg[1{+}\frac{3a^2n_{2}}{mg_2}\bigg]^{2}, \\ \label{eqn:f_1d}
        \mathcal{F}_{\rm reg}^{(1)}&{=}1{-}\frac{4}{3\pi}\sqrt{\frac{g_1a_sm}{\hbar^2}}(a_sn_{1})^{-1/2}\bigg[1+\frac{3a^2n_{1}}{mg_1}\bigg]^{3/2}.
    \end{align}
\end{subequations}
Equations \eqref{eqn:freg} introduce the three-dimensional scattering parameter $g_3=4\pi\hbar^2a_s/m$ and the energy cut-off $\Delta_{\rm cut}$ arising from the infrared divergence of the momentum-space regularization of $E_{\rm gnd}^{(2)}$. In general the scattering parameters $g_{1,2}$ differ from the three-dimensional case due to the anisotropic geometry, resulting in a confinement induced resonance \cite{olshanii_1998,salasnich_2016}. Then, the total beyond-mean-field  energy can be written in dimension $d$ using Eqs.~\eqref{eqn:freg} giving \cite{r_note}
\begin{equation}\label{eqn:etot}
    E_{\rm tot.}^{(d)}{=}\int d{\bf r}\bigg[\frac{\big|(\hat{\bf p}{-}{\bf A})\Phi\big|^2}{2m}{+}\bigg\{\frac{g_d}{2}\mathcal{F}_{\rm reg}^{(d)}{-}{\bf a}\cdot{\bf u}\bigg\}|\Phi|^4\bigg],
\end{equation}
where ${\bf A}\equiv {\bf a}|\Phi|^2$, and Eq.~\eqref{eqn:etot} is obtained by transforming the classical field equation associated with Eq.~\eqref{eqn:hem} into the moving frame using the ansatz $\psi({\bf r},t)=\Phi({\bf r}-{\bf u}t)\exp(i[m{\bf u}\cdot {\bf r}-t(mu^2/2+\mu_{d})]/\hbar)$ \cite{harikumar_1998}.

{\it Quantum droplets. } The purely mean-field model obtained from Eqs.~\eqref{eqn:hem} and \eqref{eqn:co} by taking the symmetry-breaking average $\hat{\psi}\rightarrow\langle\hat{\psi}\rangle$ already possesses an unusual nonlinear structure stemming from the nonlinear gauge potential $\langle \hat{A}({\bf r})\rangle$. It is instructive to understand when this model manifests solutions in the $({\bf a},{\bf u},g_d>0)$ parameter space. In what follows we work in the homogeneous limit which can be associated with a parameter regime where there are a large number of atoms $N\gg1$ constituting the quantum droplet. Without the LHY correction, Eq.~\eqref{eqn:etot} will in general admit solutions possessing a stable minima in the $(E_{\rm tot.}^{(d)},n_d)$ space if $g_d/2<{\bf a}\cdot{\bf u}$ with a density obtained by minimizing the energy per particle $(E_{\rm tot.}^{(d)}/L^d)/n_d$ \cite{petrov_2016}, giving $n_d=m({\bf a}\cdot{\bf u}-g_{d}/2)/a^2$. Then the mean-field chemical potential is found as
\begin{equation}\label{eqn:mumf}
    \mu_{\rm MF}=-\frac{m}{8a^2}(g_d-2{\bf a}\cdot{\bf u})^2.    
\end{equation}
In the analysis that follows pertaining to the beyond-mean-field limit, we consider the purely gauge coupled system in the absence of the background scattering length $g_d$ \cite{gnote}. Then at the mean-field level stable minima in $E_{\rm tot.}^{(d)}$ with $n_{d}>0$ only exist when ${\rm sgn}({\bf a})={\rm sgn}({\bf u})$, a situation which changes when beyond-mean-field corrections are included. In what follows we take ${\bf a}=a_{\rm d}\hat{e}_x$ similar to the experimental works in the continuum \cite{yao_2022,frolian_2022}.   
\begin{figure}[t]
    \centering
    \includegraphics[width=1\columnwidth]{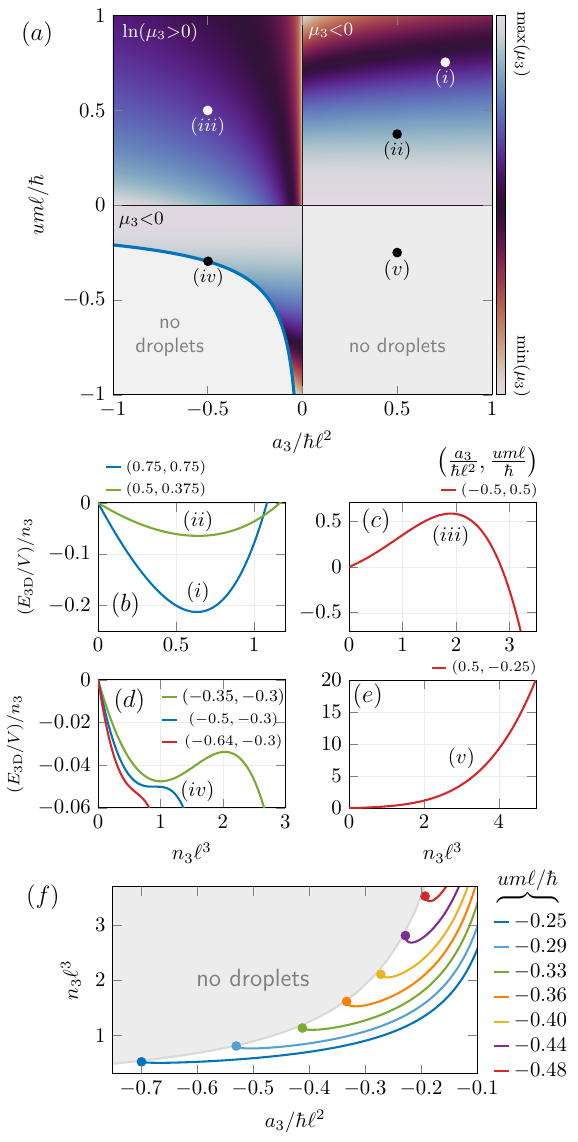}
    \caption{(color online) Three-dimensional quantum droplet phase diagram. (a) shows regions of stability obtained from the LHY energy Eqs.~\eqref{eqn:f_3d} and \eqref{eqn:etot}, with corresponding examples of the energy per particle for specific parameters given in panels (b)-(e). Examples of the allowed density $n_{3D}\ell^3$ are shown in (f) corresponding to $({\bf a}{<}0,{\bf u}{<}0)$.}
    \label{fig:h3d}
\end{figure}

Figure \ref{fig:h3d} scrutinizes the allowed three-dimensional droplet solutions in the $({\bf a},{\bf u})$ parameter space. Panel (a) shows heat maps of the beyond-mean-field chemical potential $\mu_3$ for all combinations of the signs of ${\bf a}$ and ${\bf u}$. For $({\bf a}{>}0,{\bf u}{>}0)$ the energy per particle $(E_{\rm tot.}^{(3)}/L^3)/n_3$ possesses a single minima at finite positive $n_{3}$; examples of $(E_{\rm tot.}^{(3)}/L^3)/n_3$ for this quadrant are shown in (b)(i) and (b)(ii). When $({\bf a}{<}0,{\bf u}{>}0)$ the extrema of the energy per particle are given by unstable maxima with positive chemical potential $\mu_3$, here (c)(iii) shows an example of this situation. If $({\bf a}{<}0,{\bf u}{<}0)$, quantum droplet states exist in a metastable regime where $(E_{\rm tot.}^{(3)}/L^3)/n_3$ comprises a maxima and minima whose depth depends on the particular choice of ${\bf a}$ and ${\bf u}$. This quadrant manifests points of inflection that comprise a border between the metastable droplets and a region where there are no droplets due to the energy per particle decreasing monotonically as shown in (d)(iv). The borderline between the stable and unstable region can be computed analytically as $(5\pi^2/(162\sqrt{3})|{\bf a}|=(m^2/\hbar^3)|{\bf a}\cdot{\bf u}|^2$. This means that the effect of including beyond-mean-field corrections \textit{reduces} the stable region of the $({\bf a},{\bf u})$ parameter space in this quadrant compared to the pure mean-field model Eq.~\eqref{eqn:mumf}. Meanwhile the energy per particle for the final quadrant of (a) corresponding to $({\bf a}{>}0,{\bf u}{<}0)$ does not possess any stable minima, instead $(E_{\rm tot.}^{(3)}/L^3)/n_3$ monotonically increases as shown in (e)(v). The final panel of Fig.~\ref{fig:h3d}, (f) explores how the equilibrium density $n_{3}\ell^3$ behaves as a function of the gauge potential strength $a_3/\hbar\ell^2$ in the $({\bf a}{<}0,{\bf u}{<}0)$ quadrant. For a fixed value of ${\bf u}$, there exists a value of ${\bf a}$ at which the solution terminates (see individual coloured data sets). The equilibrium density $n_{3}$ is found to increase for a given value of ${\bf u}$, with a sharp increase occurring as ${\bf a}\rightarrow 0$.  
\begin{figure}[t]
    \centering
    \includegraphics[width=1\columnwidth]{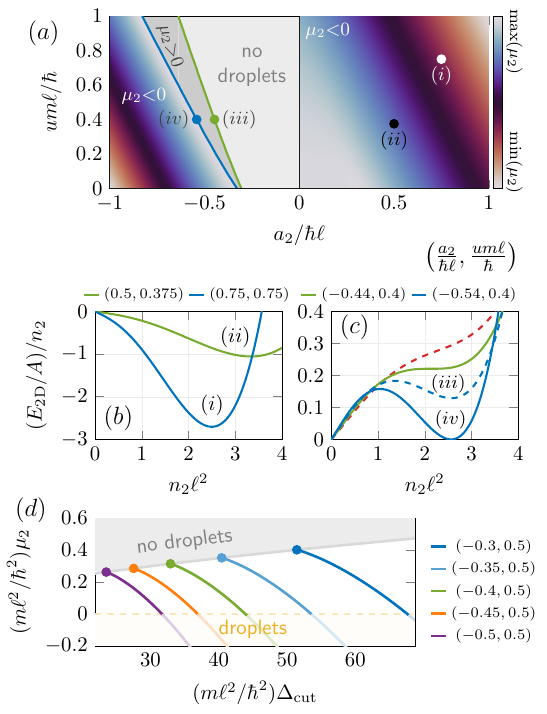}
    \caption{(color online) Two-dimensional quantum droplet phase diagram. (a) depicts heat maps of the allowed droplet solutions in the $({\bf a},{\bf u})$ parameter space with example energy per particle curves shown in (b) and (c). The dependence of the chemical potential $\mu_2$ on the cut-off energy $\Delta_{\rm cut}$ is depicted in panel (d). The numbers in brackets correspond to the values $\left(a_2/\hbar\ell,um\ell/\hbar\right)$} 
    \label{fig:h2d}
\end{figure}

Next we examine the allowed droplet solutions in two-dimensions corresponding to Eqs.~\eqref{eqn:f_2d} and \eqref{eqn:etot}, shown in Fig.~\ref{fig:h2d}. The two-dimensional regularization procedure is known to be infrared divergent, leading to the existence of a cut-off energy $\Delta_{\rm cut}$ \cite{salasnich_2016}. Choices of the cut-off depend, in general on the low-energy scattering properties of the underlying model and are generally independent of the density of the gas. Here, we work with a fixed value of $(m\ell^2/\hbar^2)\Delta_{\rm cut}=25$ in Figs.~\ref{fig:h2d}(a)-(c). We have checked that altering the cut-off modifies our results only quantitatively, the specific choice here allowing a clear visual depiction of the key physical behaviours of the parameter space. Then, regions possessing droplet solutions are shown in panel (a) in the $({\bf a},{\bf u})$ parameter space. Note that the two-dimensional energy per particle $(E_{\rm tot.}^{(2)}/L^2)/n_2$ possesses the mirror symmetry $({\bf a},{\bf u})=({-}{\bf a},{-}{\bf u})$ hence only states with ${\bf u}{>}0$ are depicted. 

When $({\bf a}{>}0,{\bf u}{>}0)$ or $({\bf a}{<}0,{\bf u}{<}0)$ single minima are found with $\mu_{2}<0$, see (b)(i) and (ii) for examples. The depth of the minima is found to increase when the strength of the gauge potential and velocity are simultaneously increased. Then for $({\bf a}{<}0,{\bf u}{>}0)$ or $({\bf a}{>}0,{\bf u}{<}0)$ the allowed solutions are separated into three regions. The energy per particle $(E_{\rm gnd.}^{(2)}/L^2)/n_2$ changes from being monotonically increasing (light grey region) to showing an inflection point (solid green line) to finally forming a metastable minima (solid blue line), each of which is depicted in (c). The dark grey region sandwiched between the inflection and metastable minima also possesses stable solutions -- however the chemical potential $\mu_{2}>0$ here, which we do not associate with a bound droplet solution (dashed blue curve in (c)). The critical density $n_{2}^{\rm crit.}$ at which the inflection point manifests can be shown to be
\begin{equation}\label{eqn:n2c}
    n_{2}^{\rm crit.}=\frac{\pi}{27}\bigg(\frac{\hbar}{\bf a}\bigg)^2\bigg[1+\sqrt{1-\frac{108}{\pi}\frac{m{\bf a}\cdot{\bf u}}{\hbar^2}}\bigg],
\end{equation}
which is independent of the cut-off $\Delta_{\rm cut}$. In contrast to the two-dimensional mean-field result Eq.~\eqref{eqn:mumf}, the beyond-mean-field model can exhibit solutions in the attractive regime when $({\bf a}{<0},{\bf u}{>}0)$. Finally panel (d) shows how the chemical potential $\mu_2$ depends on $\Delta_{\rm cut}$. The light grey shaded region corresponds to the red dashed curve in (c), while the border to the inflection region corresponds to the solid green in (a) and (c). The effect of changing $\Delta_{\rm cut}$ is explored in (d). Here each of the coloured curves shows how $\mu_2$ varies for fixed $um\ell/\hbar=\sfrac{1}{2}$ as the strength of the two-dimensional gauge potential $a_2/\hbar\ell$ is changed. Each curve starting at an inflection point (coloured circles) decreases towards $\mu_{2}=0$ (dashed gold line) where bound two-dimensional droplets are predicted. 

Low dimensional systems have provided ample opportunity for the exploration of quantum droplet phases, in particular one-dimensional models are often solvable, allowing deeper theoretical insight into the liquid-like state \cite{bottcher_2020,luo_2021,khan_2022}. Using Eqs.~\eqref{eqn:hem} and \eqref{eqn:f_1d} along with the local density approximation, an extended Gross-Pitaevskii equation can be obtained as 
\begin{equation}\label{eqn:1d_gpe}
    i\hbar\frac{\partial\psi}{\partial t}=\bigg[\frac{1}{2m}\big(\hat{\bf p}_x-{\bf A}_1\big)^2+a_1\mathcal{J}_1-\frac{2\sqrt{27}}{\pi}\frac{a_{1}^{3}}{m\hbar}|\psi|^4\bigg]\psi,
\end{equation}
where the current operator is given by $\mathcal{J}_1(x,t)\equiv(\hbar/m)\text{Im}[\psi(\partial_x+iA_1(x)/\hbar)\psi^{*}]$ and $A_1=a_1|\psi|^2$. The gauge potential appearing in Eq.~\eqref{eqn:1d_gpe} can be decoupled using the Jordan-Wigner-like transformation \cite{aglietti_1996}
\begin{equation}
    \psi(x,t)=\exp\bigg(\frac{ia_1}{\hbar}\int\limits^{x}_{-\infty}dx'|\Phi(x',t)|^2\bigg)\Phi(x,t)
\end{equation}
which gives a current-coupled derivative cubic-quintic Schr\"odinger equation
\begin{equation}\label{eqn:dcq_gpe}
    i\hbar\frac{\partial\Phi}{\partial t}=\bigg[-\frac{\hbar^2}{2m}\frac{\partial^2}{\partial x^2}-2a_1j(x,t)+\frac{2\sqrt{27}}{\pi}\frac{a_{1}^{3}}{m\hbar}|\Phi|^4\bigg]\Phi,
\end{equation}
along with the transformed current nonlinearity $j(x,t)=(\hbar/m)\text{Im}[\Phi\partial_x\Phi^{*}]$. Applying the transformation $\Phi(x,t)=\xi(x-ut)\exp(i[umx-t(mu^2/2+\mu)]/\hbar)$ to Eq.~\eqref{eqn:dcq_gpe} leads to the cubic quintic Schr\"odinger model $\mu\xi=-(\hbar^2/2m)\partial_{x}^{2}\xi-2a_1u\xi^3+(a_{1}^{3}/m\hbar)(2\sqrt{27}/\pi)\xi^5$. In writing Eq.~\eqref{eqn:dcq_gpe} we have chosen the gauge potential strength such that $a_1\rightarrow-a_1$ motivated by the known solutions studied in nonlinear optics and strongly-interacting Bose gases \cite{birnbaum_2008,kolomeisky_2000}. Two classes of stable solutions exist, the first with vanishing boundary conditions $\lim_{x\rightarrow\pm\infty}\xi(x-ut)=0$ gives rise to the quantum droplet solution
\begin{equation}\label{eqn:1d_qd}
    \frac{\Phi(x,t)}{\sqrt{n_{1}}}{=}\frac{\sqrt{\mu/\mu_{1}}\exp\big(i[umx{-}t(mu^2/2{+}\mu)]/\hbar\big)}{\sqrt{1{+}\sqrt{1{-}\mu/\mu_{1}}\cosh\big(\sqrt{-8m\mu}(x{-}ut)/\hbar\big)}}.
\end{equation}
\begin{figure}[t]
    \centering
    \includegraphics[width=0.9\columnwidth]{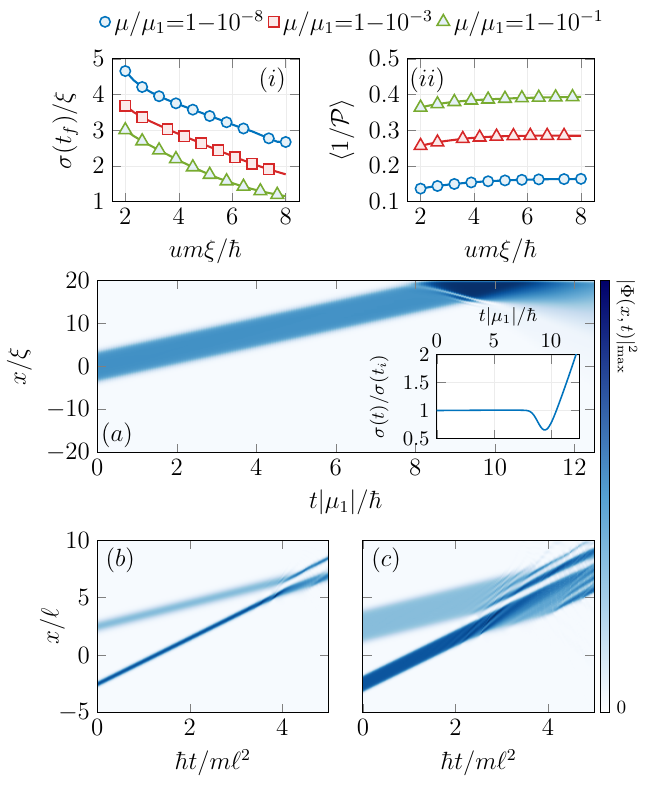}
    \caption{(color online) Chiral droplet dynamics. (a)-(c) shows the dynamics of single (a) and pairs (b,c) of chiral droplets (see Eq.~\eqref{eqn:dcq_gpe} and \eqref{eqn:1d_qd}). Panels (i) and (ii) compute the width $\sigma(t_f)/\xi$ and time-averaged inverse participation ratio $\langle1/\mathcal{P}\rangle$, Eq.~\eqref{eqn:ipr} as a function of $um\xi/\hbar$.}
    \label{fig:1d}
\end{figure}
Analogous to the situation in higher dimensions the droplet solution in Eq.~\eqref{eqn:1d_qd} possess an equilibrium density $n_{1}=(\pi/4\sqrt{3})um\hbar/a_{1}^{2}$ with associated chemical potential $\mu_{1}=-a_1un_{1}/2$, while the total atom number can be found using $N(\mu)=\int^{\infty}_{-\infty} dx\ |\Phi(x,t)|^2$ as
\begin{equation}\label{eqn:1d_anum}
    N(\mu)=\sqrt{2}n_{1}\xi_{1}\text{artanh}\bigg(\frac{1-\sqrt{1-\mu/\mu_{1}}}{\sqrt{\mu/\mu_{1}}}\bigg),
\end{equation}
with the length scale $\xi_1=\hbar/\sqrt{-m\mu_1}$. We note that the atom number $N(\mu)$ in Eq.~\eqref{eqn:1d_anum} diverges as $\mu\rightarrow\mu_{1}$. Next we compute the surface tension $\sigma_1$ of the quantum droplet in Eq.~\eqref{eqn:1d_qd} defined as $\sigma_1=\int^{\infty}_{-\infty}dx\big[\varepsilon(x)-\mu_1|\Phi(x)|^2\big]$ \cite{bulgac_2002}, where $\varepsilon(x)=(\hbar^2/2m)|\partial_x\Phi|^2+(2\sqrt{27}/3\pi)(a_{1}^{3}/m\hbar)|\Phi|^6$, which results in 
\begin{align}\label{eqn:st}
    \frac{\sigma_{1}}{n_1\xi_1}=\frac{1}{\sqrt{2}}\bigg(\big(mu^2+4\mu_1\big)\frac{N(\mu)}{\sqrt{2}n_1\xi_1}{-}\sqrt{\mu_1\mu}\bigg).
\end{align}
Through the equilibrium chemical potential $\mu_1$ the surface tension depends in general on both the strength $a_1$ of the gauge potential and unusually the velocity $u$ of the quantum droplet. The surface tension $\sigma_1$ 
will for fixed $a_1$ increase monotonically with $u$, while for fixed $u$ will instead decrease with increasing $a_1$. 

The corresponding dark soliton-like solution is instead given by 
\begin{align}\nonumber
    \frac{\Phi_{\rm d}(x,t)}{\sqrt{\rho_1}}=&\sinh{\big(k(\mu)(x-ut)/\xi_1\big)}\\&\times\frac{\exp\big(i[umx{-}t(mu^2/2{+}\mu)]/\hbar\big)}{\sqrt{\chi(\mu)+\sinh^2{\big(k(\mu)(x-ut)/\xi_1\big)}}},\label{eqn:dqd}
\end{align}
with the asymptotic (vacua) appearing in Eq.~\eqref{eqn:dqd} defined as $\rho_1(\mu)=\frac{2}{3}\big(1+\sqrt{1-3\mu/4\mu_1}\big)$. Then the function $\chi(\mu)$ is given by
\begin{equation}
    \chi(\mu)=\frac{3\big(3\mu/4\mu_1-1-\sqrt{1-3\mu/4\mu_1}\big)}{3\mu/2\mu_1-1-\sqrt{1-3\mu/4\mu_1}},
\end{equation}
while the dimensionless function $k(\mu)$ appearing in the definition of $\Phi_{\rm d}(x,t)$ in Eq.~\eqref{eqn:dqd}  is given by $k(\mu)=2\sqrt{3}\sqrt{1-3\mu/4\mu_1+\sqrt{1-3\mu/4\mu_1}}$. Equation \eqref{eqn:dqd} describes a propagating solitary wave whose amplitude $\Phi_{\rm d}(x,t)$ develops a low-density region centred around the spatial origin that increases in size as $\mu\rightarrow\mu_1$. A regularized atom number \cite{kivshar_1998} can be associated with Eq.~\eqref{eqn:dqd} defined as $N_{\rm d}(\mu)=\int dx(\rho_1-|\Phi_{\rm d}(x,t)|^2)$, given by
\begin{equation}
    N_{\rm d}(\mu)=\frac{2\rho_{1}(\mu)}{k(\mu)}\sqrt{\frac{\chi(\mu)}{\chi(\mu){-}1}}\text{artanh}\bigg[\sqrt{\frac{\chi(\mu){-}1}{\chi(\mu)}}\bigg].
\end{equation}
The dynamics of single and pairs of chiral droplets are explored in Fig.~\ref{fig:1d}. The dynamics of a single quantum droplet solution to Eq.~\eqref{eqn:dcq_gpe} with $um\xi/\hbar=2$ and $\mu/\mu_1=1-10^{-8}$ is shown in (a). Here the droplet propagates before expanding due to collision with the edge of the numerical box. The inset computes the width $\sigma(t)/\sigma(t_i)$ as a function of time. To further quantify the dynamics, the droplet's width $\sigma(f_f)/\xi$ as a function of velocity $um\xi/\hbar$ and time-averaged inverse participation ratio (IPR) $\langle1/\mathcal{P}\rangle=(1/T)\int_{0}^{T}dt/\mathcal{P}(t)$ with 
\begin{equation}\label{eqn:ipr}
    \frac{1}{\mathcal{P}(t)}=\frac{\int dx\ n(x,t)^2}{(\int dx\ n(x,t))^2}
\end{equation} 
are shown in panels (i) and (ii) respectively for three fixed values of $\mu/\mu_1$. Smaller droplets (green triangles) correspond to the smallest values of $\sigma(t_f)$ and the largest values of $\langle1/\mathcal{P}\rangle$. The IPR quantifies how localized a solution is.  The larger droplets possess smaller values of $\langle1/\mathcal{P}\rangle$ due to the increased dominance of the beyond-mean-field term in Eq.~\eqref{eqn:dcq_gpe}. Panels (b) and (c) show the dynamics of a pair of droplets with $a_1/\hbar=(3\pi/2\sqrt{27})^{1/3}$ and $u_{1,2}m\ell/\hbar=(2,1)$, with droplet normalizations $N=2$ (a) and $N=7$ (b). Increasing the atom number has the effect of changing the dynamics from particle-like (b) to showing non-integrable effects, where indeed the collision in (c) shows chiral droplet fusion. The initial phase difference between the droplets is $\delta=\pi$ in both cases. 

Finally we can assess the viability of the theoretical predictions obtained in this work with experimental values. In the work of Fr\"olian et al.,\cite{frolian_2022}, the current strength was given by 
\begin{equation}\label{eqn:a1d}
    a_1=\frac{m_{\rm eff}k_{\rm R}(g_{\uparrow\uparrow}-g_{\downarrow\downarrow})}{2\pi a_{r}^{2}\Omega}
\end{equation}
where $1/m_{\rm eff}=1-4E_{\rm R}/\hbar\Omega$ defines the effective mass of the atoms, $k_{\rm R}$ is the recoil momentum, $a_r$ is the radial length scale, $g_{ij}=4\pi\hbar^2a_{ij}/m_0$ encapsulates the $s$-wave scattering lengths with $a_{\uparrow\uparrow}=-4.9a_0$, $a_{\downarrow\downarrow}=24.6a_0$, and $\Omega$ is the two-photon Rabi coupling strength. Using values appropriate for $^{39}$K and a radial trap strength $\omega_r=2\pi\times10 {\rm kHz}$ \cite{haller_2009}, we obtain $a_{1}\simeq -6\times 10^{-36}$Js which along with a typical velocity $u= 17 {\rm mm s^{-1}}$ gives a feasible one-dimensional equilibrium density of $n_1=4\times 10^{9}{\rm m}^{-1}$. 

{\it Outlook. } We derived the beyond-mean-field (LHY) corrections appropriate to a density-dependent gauge theory which are free from the issues associated with the complex-valued beyond-mean-field ground state energy inherent to theoretical treatments of Bose-Bose and dipolar condensates \cite{cikojevi_2020,hu_2025}. We explored the dimensional-dependence of the droplet solutions, scrutinising the allowed stable equilibrium phases in the full two- and three-dimensional parameter space. Moving quantum droplet solutions in this model are supported by a single interaction parameter, and exist in regimes where the mean-field model is unstable in the two-dimensional case. In one dimension exact analytical solutions were obtained, along with an expression for the surface tension of the droplet. Numerical simulations explored the nonlinear dynamics of the beyond-mean-field model, revealing the fundamental phenomenology of the chiral droplet states.

The beyond-mean-field system explored in this work offers several opportunities for future work. Lower dimensional systems provide a wealth of phenomena with potential applications to quantum technologies like atomtronics \cite{amico_2021}. The allowed superfluid behaviour would also constitute an interesting avenue to explore, particularly the case for vortex states in two dimensions \cite{tengstrand_2019} where an additional rigid-body rotation would give insight into the topological physics of the model.    
 
{\it Acknowledgments. } We thank Antonino Flachi, Stewart Lang, Antonio Mu\~noz Mateo, Joel Priestley, Gerard Valent\'{\i}-Rojas and Ewan Wright for helpful discussions. This research was supported by the Australian Research Council Centre of Excellence in Future Low-Energy Electronics Technologies (Project No. CE170100039) and funded by the Australian government, and by the Japan Society of Promotion of Science Grant-in-Aid for Scientific Research (KAKENHI Grant No. JP20K14376).


\begin{thebibliography}{99}
%
\bibitem{bruus_book}
\bref{H. Bruus and K. Flensberg, {\it Quantum Theory in Condensed Matter Physics} }{https://academic.oup.com/book/54902}{(Oxford University Press, 2006).}
%
\bibitem{klitzing_2020}
\bref{K. von Klitzing, T. Chackraborty, P. Kim, V. Madhaven, X. Dai, J. McIver, Y. Tokura, L. Savary, D. Smirnova, A. Maria Rey, C. Felser, J. Gooth, and X. Qi, }{https://doi.org/10.1038/s42254-020-0209-1}{Nat. Rev. Phys. 2, 397 (2020).}
%
\bibitem{franchini_book}
\bref{F. Franchini, {\it An Introduction to Integrable Techniques for One-Dimensional Quantum Systems} }{https://doi.org/10.1007/978-3-319-48487-7}{(Springer International Publishing, Cham, 2017).}
%
\bibitem{cazalilla_2011}
\bref{M. A. Cazalilla, R. Citro, T. Giamarchi, E. Orignac, and M. Rigol, }{https://doi.org/10.1103/RevModPhys.83.1405}{Rev. Mod. Phys. 83, 1405 (2011).}
%
\bibitem{sowinski_2019}
\bref{Tomasz Sowi\'nski and Miguel \'Angel Garc\'ia-March, }{https://iopscience.iop.org/article/10.1088/1361-6633/ab3a80}{Rep. Prog. Phys. 82, 104401 (2019).}
%
\bibitem{mistakidis_2023}
\bref{S. I. Mistakidis, A. G. Volosniev, R. E. Barfknecht, T. Fogarty, Th. Busch, A. Foerster, P. Schmelcher, and N. T. Zinner, }{https://doi.org/10.1016/j.physrep.2023.10.004}{Phys. Rep. 1042, 1 (2023).}
%
\bibitem{ohashi_2020}
\bref{Y. Ohashi, H. Tajima, and P. van Wyk, }{https://doi.org/10.1016/j.ppnp.2019.103739}{Prog. Part. Nucl. Phys. 111, 103739 (2020).}
%
\bibitem{strinati_2018}
\bref{G. C. Strinati, P. Pieri, G. R\"opke, P. Schuck, and M. Urban, }{https://doi.org/10.1016/j.physrep.2018.02.004}{Phys. Rep. 738, 1 (2018).}
%
\bibitem{kadau_2016}
\bref{H. Kadau, M. Schmitt, M. Wenzel, C. Wink, T. Maier, I. F.-Barbut, and T. Pfau, }{https://doi.org/10.1038/nature16485}{Nature 530, 194 (2016).}
%
\bibitem{schmitt_2016}
\bref{M. Schmitt, M. Wenzel, F. B\"ottcher, I. F.-Barbut, and T. Pfau, }{https://doi.org/10.1038/nature20126}{Nature 539, 259 (2016).}
%
\bibitem{barbut_2016}
\bref{I. F.-Barbut, H. Kadau, M. Schmitt, M. Wenzel, and T. Pfau, }{https://doi.org/10.1103/PhysRevLett.116.215301}{Phys. Rev. Lett. 116, 215301 (2016).}
%
\bibitem{cabrera_2018}
\bref{C. R. Cabrera, L. Tanzi, J. Sanz, B. Naylor, P. Thomas, P. Cheiney, and L. Tarruell, }{https://doi.org/10.1126/science.aao5686}{Science 359, 301 (2018).}
%
\bibitem{semeghini_2018}
\bref{G. Semeghini, G. Ferioli, L. Masi, C. Mazzinghi, L. Wolswijk, F. Minardi, M. Modugno, G. Modugno, M. Inguscio, and M. Fattori, }{https://doi.org/10.1103/PhysRevLett.120.235301}{Phys. Rev. Lett. 120, 235301 (2018).}
%
\bibitem{cheiney_2018}
\bref{P. Cheiney, C. R. Cabrera, J. Sanz, B. Naylor, L. Tanzi, and L. Tarruell, }{https://doi.org/10.1103/PhysRevLett.120.135301}{Phys. Rev. Lett. 120, 135301 (2018).}
%
\bibitem{ferioli_2019}
\bref{G. Ferioli, G. Semeghini, L. Masi, G. Giusti, G. Modugno, M. Inguscio, A. Gallem\'i, A. Recati, and M. Fattori, }{https://doi.org/10.1103/PhysRevLett.122.090401}{Phys. Rev. Lett. 122, 090401 (2019).}
%
\bibitem{derrico_2019}
\bref{C. D'Errico, A. Burchianti, M. Prevedelli, L. Salasnich, F. Ancilotto, M. Modugno, F. Minardi, and C. Fort, }{https://doi.org/10.1103/PhysRevResearch.1.033155}{Phys. Rev. Research 1, 033155 (2019).}
%
\bibitem{cavicchioli_2025}
\bref{L. Cavicchioli, C. Fort, F. Ancilotto, M. Modugno, F. Minardi, A. Burchianti, }{https://doi.org/10.1103/PhysRevLett.134.093401}{Phys. Rev. Lett. 134, 093401 (2025).}
%
\bibitem{lima_2011}
\bref{A. R. P. Lima and A. Pelster, }{https://doi.org/10.1103/PhysRevA.84.041604}{Phys. Rev. A 84, 041604(R) (2011).}
%
\bibitem{lima_2012}
\bref{A. R. P. Lima and A. Pelster, }{https://doi.org/10.1103/PhysRevA.86.063609}{Phys. Rev. A 86, 063609 (2012).}
%
\bibitem{petrov_2015}
\bref{D. S. Petrov, }{https://doi.org/10.1103/PhysRevLett.115.155302}{Phys. Rev. Lett. 115, 155302 (2015).}
%
\bibitem{lee_1957}
\bref{T. D. Lee, K. Huang, and C. N. Yang, }{https://doi.org/10.1103/PhysRev.106.1135}{Phys. Rev. 106, 1135 (1957).}
%
\bibitem{jorgensen_2018}
\bref{N. B. J\o rgensen, G. M. Bruun, and J. J. Arlt, }{https://doi.org/10.1103/PhysRevLett.121.173403}{Phys. Rev. Lett. 121, 173403 (2018).}
%
\bibitem{skov_2021}
\bref{T. G. Skov, Magnus G. Skou, N. B. J\o rgensen, and J. J. Arlt, }{https://doi.org/10.1103/PhysRevLett.126.230404}{Phys. Rev. Lett. 126, 230404 (2021).}
%
\bibitem{chiquillo_2019}
\bref{E. Chiquillo, }{https://doi.org/10.1103/PhysRevA.99.051601}{Phys. Rev. A 99, 051601(R) (2019).}
%
\bibitem{uchino_2010}
\bref{S. Uchino, M. Kobayashi, and M. Ueda, }{https://doi.org/10.1103/PhysRevA.81.063632}{Phys. Rev. A 81, 063632 (2010).}
%
\bibitem{yogurt_2023}
\bref{T. A. Yo\u gurt, A. Kele\c s, and M. \"O. Oktel, }{https://doi.org/10.1103/PhysRevA.107.023322}{Phys. Rev. A 107, 023322 (2023).}
%
\bibitem{adhikari_2018}
\bref{S. Adhikari, }{https://doi.org/10.1088/1612-202X/aacb0a}{Laser Phys. Lett. 15, 095501 (2018).}
%
\bibitem{rakshit_2019}
\bref{D. Rakshit, T. Karpiuk, M. Brewczyk, and M. Gajda, }{https://doi.org/10.21468/SciPostPhys.6.6.079}{SciPost Phys. 6, 079 (2019).}
%
\bibitem{edler_2017}
\bref{D. Edler, C. Mishra, F. W\"achtler, R. Nath, S. Sinha, and L. Santos, 
}{https://doi.org/10.1103/PhysRevLett.119.050403}{Phys. Rev. Lett. 119, 050403 (2017).}
%
\bibitem{tanzi_2019}
\bref{L. Tanzi, E. Lucioni, F. Fam\`a, J. Catani, A. Fioretti, C. Gabbanini, R. N. Bisset, L. Santos, and G. Modugno, }{https://link.aps.org/doi/10.1103/PhysRevLett.122.130405}{Phys. Rev. Lett. 122, 130405 (2019).}
%
\bibitem{bottcher_2019}
\bref{F. B\"ottcher, J.-N. Schmidt, M. Wenzel, J. Hertkorn, M. Guo, T. Langen, and T. Pfau, }{https://doi.org/10.1103/PhysRevX.9.011051}{Phys. Rev. X 9, 011051 (2019).}
%
\bibitem{chomaz_2019}
\bref{L. Chomaz, D. Petter, P. Ilzh\"ofer, G. Natale, A. Trautmann, C. Politi, G. Durastante, R. M. W. van Bijnen, A. Patscheider, M. Sohmen, M. J. Mark, and F. Ferlaino, }{https://doi.org/10.1103/PhysRevX.9.021012}{Phys. Rev. X 9, 021012 (2019).}
%
\bibitem{barranco_2006}
\bref{M. Barranco, R. Guardiola, S. Hern\'andez, R. Mayol, J. Navarro, and M. Pi, }{https://doi.org/10.1007/s10909-005-9267-0}{J. Low Temp. Phys. 142, 1 (2006).}
%
\bibitem{walker_2022}
\bref{J. G. M. Walker, G. R. M. Robb, G.-L. Oppo, and T. Ackemann, }{https://doi.org/10.1103/PhysRevA.105.063305}{Phys. Rev. A 105, 063305 (2022).}
%
\bibitem{hunter_2014}
\bref{A. E. Almand-Hunter, H. Li, S. T. Cundiff, M. Mootz, M. Kira and S. W. Koch, }{https://doi.org/10.1038/nature12994}{Nature 506, 471 (2014).}
%
\bibitem{wilson_2018}
\bref{K. E. Wilson, N. Westerberg, M. Valiente, C. W. Duncan, E. M. Wright, P. \"Ohberg, and D. Faccio, }{https://doi.org/10.1103/PhysRevLett.121.133903}{Phys. Rev. Lett. 121, 133903 (2018).}
%
\bibitem{schafer_2020}
\bref{F. Sch\"afer, T. Fukuhara, S. Sugawa, Y. Takasu, and Y. Takahashi, }{https://doi.org/10.1038/s42254-020-0195-3}{Nat. Rev. Phys. 2, 411 (2020).}
%
\bibitem{lin_2009}
\bref{Y.-J. Lin, R. L. Compton, K. Jim\'enez-Garc\'ia, J. V. Porto, and  I. B. Spielman, }{https://doi.org/10.1038/nature08609}{Nature 462, 628 (2009).}
%
\bibitem{lin_2009b}
\bref{Y.-J. Lin, R. L. Compton, A. R. Perry, W. D. Phillips, J. V. Porto, and I. B. Spielman, }{https://doi.org/10.1103/PhysRevLett.102.130401}{Phys. Rev. Lett. 102, 130401 (2009).}
%
\bibitem{lin_2011}
\bref{Y.-Ju Lin, R. L. Compton, K. Jim\'enez-Garc\'ia, W. D. Phillips, J. V. Porto, and I. B. Spielman, }{https://doi.org/10.1038/nphys1954}{Nat. Phys. 7, 531 (2011).}
%
\bibitem{lin_2011b}
\bref{Y.-J. Lin, K. Jim\'enez-Garc\'ia, and I. B Spielman, }{https://doi.org/10.1038/nature09887}{Nature 471, 83 (2011).}
%
\bibitem{beeler_2013}
\bref{M. C. Beeler, R. A. Williams, K. Jim\'enez-Garc\'ia, L. J. LeBlanc, A. R. Perry, I. B. Spielman, }{https://doi.org/10.1038/nature12185}{Nature 498, 201 (2013).}
%
\bibitem{celi_2014}
\bref{A. Celi, P. Massignan, J. Ruseckas, N. Goldman, I. B. Spielman, G. Juzeli\=unas, and M. Lewenstein, }{https://doi.org/10.1103/PhysRevLett.112.043001}{Phys. Rev. Lett. 112, 043001 (2014).}
%
\bibitem{clark_2018}
\bref{L. W. Clark, B. M. Anderson, L. Feng, A. Gaj, K. Levin, and C. Chin, }{https://doi.org/10.1103/PhysRevLett.121.030402}{Phys. Rev. Lett. 121, 030402 (2018).}
%
\bibitem{kwan_2024}
\bref{J. Kwan, P. Segura, Y. Li, S. Kim, A. V. Gorshkov, A. Eckardt, B. Bakkali-Hassani, and M. Greiner, }
{https://doi.org/10.1126/science.adi3252}{Science 386, 1055 (2024).}
%
\bibitem{gorg_2019}
\bref{F. G\"org, K. Sandholzer, J. Minguzzi, R. Desbuquois, M. Messer, and T. Esslinger, }{https://doi.org/10.1038/s41567-019-0615-4}{Nat. Phys. 15, 1161 (2019).}
%
\bibitem{lienhard_2020}
\bref{V. Lienhard, P. Scholl, S. Weber, D. Barredo, S. de L\'es\'eleuc, R. Bai, N. Lang, M. Fleischhauer, H. P. B\"uchler, T. Lahaye, and A. Browaeys, }{https://doi.org/10.1103/PhysRevX.10.021031}{Phys. Rev. X 10, 021031 (2020).}
%
\bibitem{yao_2022}
\bref{K.-X. Yao, Z. Zhang, and C. Chin, }{https://doi.org/10.1038/s41586-021-04250-3}{Nature 602, 68 (2022).}
%
\bibitem{aglietti_1996}
\bref{U. Aglietti, L. Griguolo, R. Jackiw, S.-Y. Pi, and D. Seminara, }{https://doi.org/10.1103/PhysRevLett.77.4406}{Phys. Rev. Lett. 77, 4406 (1996).}
%
\bibitem{frolian_2022}
\bref{A. Fr\"olian, C. S. Chisholm, E. Neri, C. R. Cabrera, R. Ramos, A. Celi, and L. Tarruell, }{https://doi.org/10.1038/s41586-022-04943-3}{Nature 602, 68 (2022).}
%
\bibitem{chisholm_2022}
\bref{C. S. Chisholm, A. Fr\"olian, E. Neri, R. Ramos, L. Tarruell, and A. Celi, }{https://doi.org/10.1103/PhysRevResearch.4.043088}{Phys. Rev. Research 4, 043088 (2022).}
%
\bibitem{dingwall_2019}
\bref{R. J. Dingwall and P. \"Ohberg, }{https://doi.org/10.1103/PhysRevA.99.023609}{Phys. Rev. A 99, 023609 (2019).}
%
\bibitem{bhat_2021}
\bref{I. A. Bhat, S. Sivaprakasam, and B. A. Malomed, }{https://doi.org/10.1103/PhysRevE.103.032206}{Phys. Rev. E 103, 032206 (2021).}
%
\bibitem{jia_2022}
\bref{Q. Jia, H. Qiu, and A. M. Mateo, }{https://doi.org/10.1103/PhysRevA.106.063314}{Phys. Rev. A 106, 063314 (2022).}
%
\bibitem{chen_2022}
\bref{L. Chen and Q. Zhu, }{https://doi.org/10.1088/1367-2630/ac6cfd}{New J. Phys. 24 053044, (2022).}
%
\bibitem{zhang_2023}
\bref{A.-Q. Zhang, C. Jiao, Z.-F. Yu, J. Wang, A.-X. Zhang, and J.-K. Xue, }{https://doi.org/10.1103/PhysRevE.107.024218}{Phys. Rev. E 107, 024218 (2023).}
%
\bibitem{gao_2023}
\bref{R. Gao, X. Qiao, Y.-E. Ma, Y. Jian, A.-X. Zhang and J.-K. Xue et al., }{https://doi.org/10.1209/0295-5075/acbf6e}{Europhys. Lett. 141, 55003 (2023).}
%
\bibitem{xu_2023}
\bref{J. Xu, Q. Jia, H. Qiu, and A. M. Mateo, }{https://doi.org/10.1103/PhysRevA.108.053313}{Phys. Rev. A 108, 053313 (2023).}
%
\bibitem{arazo_2023}
\bref{M. Arazo, M. Guilleumas, R. Mayol, V. Delgado, and A. M. Mateo, }{https://doi.org/10.1103/PhysRevA.108.053302}{Phys. Rev. A 108, 053302 (2023).}
%
\bibitem{faugno_2024}
\bref{W. N. Faugno, Mario Salerno, and Tomoki Ozawa, }{https://doi.org/10.1103/PhysRevLett.132.023401}{Phys. Rev. Lett. 132, 023401 (2024).}
%
\bibitem{zezyulin_2018}
\bref{D. A. Zezyulin, D. R. Gulevich, D. V. Skryabin, and I. A. Shelykh, }{https://doi.org/10.1103/PhysRevB.97.161302}{Phys. Rev. B 97, 161302(R) (2018).}
%
\bibitem{xu_2021}
\bref{P. Xu, T.-Shu Deng, W. Zheng, and H. Zhai, }{https://doi.org/10.1103/PhysRevA.103.L061302}{Phys. Rev. A 103, L061302 (2021).}
%
\bibitem{abdullaev_2023}
\bref{F. Kh. Abdullaev, M. S. A. Hadi, B. Umarov, L. A. Taib, and M. Salerno, }{https://doi.org/10.1103/PhysRevE.107.044218}{Phys. Rev. E 107, 044218 (2023).}
%
\bibitem{arazo_2024}
\bref{M. Arazo, M. Guilleumas, R. Mayol, V. Delgado, and A. M. Mateo, }{https://doi.org/10.1103/PhysRevA.110.023316}{Phys. Rev. A 110, 023316 (2024).}
%
\bibitem{bhat_2023}
\bref{I. A. Bhat, T. Mithun, and B. Dey, }{https://doi.org/10.1103/PhysRevE.107.044210}{Phys. Rev. E 107, 044210 (2023).}
%
\bibitem{bhat_2024}
\bref{I. A. Bhat and B. Dey, }{https://doi.org/10.1103/PhysRevE.110.024208}{Phys. Rev. E 110, 024208 (2024).}
%
\bibitem{gao_2024}
\bref{M.-Z. Zhou, Y.-E. Ma, S.-D. Xu, L.-L. Mi, A.-X. Zhang, and J.-K. Xue, }{https://doi.org/10.1088/1361-6455/ad41c0}{J. Phys. B: At. Mol. Opt. Phys. 57 125301 (2024).}
%
\bibitem{rojas_2020}
\bref{G. Valent\'i-Rojas, N. Westerberg, and P. \"Ohberg, }{https://doi.org/10.1103/PhysRevResearch.2.033453}{Phys. Rev. Research 2, 033453 (2020).}
%
\bibitem{rojas_2023}
\bref{G. Valent\'i-Rojas, A. J. Baker, A. Celi, and P. \"Ohberg, }{https://doi.org/10.1103/PhysRevResearch.5.023128}{Phys. Rev. Research 5, 023128 (2023).}
%
\bibitem{pethick_book}
\bref{C. J. Pethick and H. Smith, {\it Bose-Einstein Condensation in
Dilute Gases}, }{https://doi.org/10.1017/CBO9780511802850}{(Cambridge University Press, Cambridge, 2002).}
%
\bibitem{ueda_book}
\bref{M. Ueda, {\it Fundamentals and New Frontiers of Bose-Einstein Condensation}, }{https://doi.org/10.1142/7216}{(World Scientific, Singapore, 2010).}
%
\bibitem{olshanii_1998}
\bref{M. Olshanii, }{https://doi.org/10.1103/PhysRevLett.81.938}{Phys. Rev. Lett. 81, 938 (1998).}
%
\bibitem{salasnich_2016}
\bref{L. Salasnich and F. Toigo, }{https://doi.org/10.1016/j.physrep.2016.06.003}{Phys. Rep. 640, 1 (2016).}
%
\bibitem{r_note}
\bref{The term $a^2n_{d}^{3}/2m$ appearing in $E_{\rm gnd.}^{(d)}/L^d$ is not included in Eq.~\eqref{eqn:etot} so that the underlying mean-field theory agrees with existing works, Refs.~\cite{aglietti_1996,chisholm_2022}. Including this term contributes only quantitatively to $n_{d}{\bf A}^2/2m$.}{}{}
%
\bibitem{harikumar_1998}
\bref{E. Harikumar, C. Nagaraja Kumar, and M. Sivakumar, }{https://doi.org/10.1103/PhysRevD.58.107703}{Phys. Rev. D 58, 107703, (1998).}
%
\bibitem{petrov_2016}
\bref{D. S. Petrov and G. E. Astrakharchik, }{https://doi.org/10.1103/PhysRevLett.117.100401}{Phys. Rev. Lett. 117, 100401 (2016).}
%
\bibitem{gnote}
\bref{One could also include the background $s$-wave scattering term $g_{d}$, but our goal here is to understand the interplay of the gauge and beyond-mean-field terms.}{}{}
%
\bibitem{bottcher_2020}
\bref{F. B\"ottcher, J.-N. Schmidt, J. Hertkorn, K. S. H. Ng, S. D. Graham, M. Guo, T. Langen, and T. Pfau, }{https://doi.org/10.1088/1361-6633/abc9ab}{Rep. Prog. Phys. 84, 012403 (2020).}
%
\bibitem{luo_2021}
\bref{Z.-H. Luo, W. Pang, B. Liu, Y.-Y. Li, and B. A. Malomed, }{https://doi.org/10.1007/s11467-020-1020-2}{Font. Phys. 16, 32201 (2021).}
%
\bibitem{khan_2022}
\bref{A. Khan and A. Debnath, }{https://doi.org/10.3389/fphy.2022.887338 }{Front. Phys. 10, 887338 (2022).}
%
\bibitem{birnbaum_2008}
\bref{Z. Birnbaum and B. A. Malomed, }{https://doi.org/10.1016/j.physd.2008.08.005}{Physica D 237, 3252 (2008).}
%
\bibitem{kolomeisky_2000}
\bref{E. B. Kolomeisky, T. J. Newman, J. P. Straley, and X. Qi, }{https://doi.org/10.1103/PhysRevLett.85.1146}{Phys. Rev. Lett. 85, 1146 (2000).}
%
\bibitem{bulgac_2002}
\bref{A. Bulgac, }{https://doi.org/10.1103/PhysRevLett.89.050402}{Phys. Rev. Lett. 89, 050402 (2002).}
%
\bibitem{kivshar_1998}
\bref{Y. S. Kivshar and B. L.-Davies, }{https://doi.org/10.1016/S0370-1573(97)00073-2}{Phys. Rep. 298, 81 (1998).}
%
\bibitem{haller_2009}
\bref{E. Haller, M. Gustavsson, M. J. Mark, J. G. Danzl, R. Hart, G. Pupillo, and H.-C. N\"agerl, }{https://doi.org/10.1126/science.1175850}{Science 325, 1224 (2009).}
%
\bibitem{cikojevi_2020}
\bref{V. Cikojevi\'c, L. V. Marki\'c, and J. Boronat, }{https://doi.org/10.1088/1367-2630/ab867a}{New J. Phys. 22 053045 (2020).}
%
\bibitem{hu_2025}
\bref{H. Hu, J. Wang, H. Pu, and X.-J. Liu, }{ https://doi.org/10.1103/PhysRevA.111.023309}{Phys. Rev. A 111, 023309 (2025).}
%
\bibitem{amico_2021}
\bref{L. Amico et al., }{https://doi.org/10.1116/5.0026178}{AVS Quantum Sci. 3, 039201 (2021).}
%
\bibitem{tengstrand_2019}
\bref{M. N. Tengstrand, P. St\"urmer, E. \"O. Karabulut, and S. M.
Reimann, }{https://doi.org/10.1103/PhysRevLett.123.160405}{Phys. Rev. Lett. 123, 160405 (2019).}
%
\end{thebibliography}
\end{document}